# Optically Controlled Polarization in Highly Oriented Ferroelectric Thin Films

Hitesh Borkar[1,2], M Tomar[3], Vinay Gupta[4], Ram S. Katiyar[5], J. F. Scott[6], Ashok Kumar[1,2,*]

[1]CSIR-National Physical Laboratory, Dr. K. S. Krishnan Marg, New Delhi 110012, India
[2]Academy of Scientific and Innovative Research (AcSIR), CSIR-National Physical Laboratory (CSIR-NPL) Campus, Dr. K. S. Krishnan Road, New Delhi 110012, India
[3]Department of Physics, Miranda House, University of Delhi, Delhi 110007, India
[4]Department of Physics and Astrophysics, University of Delhi, Delhi 110007, India
[5]Department of Physics and Institute for Functional Nanomaterials, University of Puerto Rico, San Juan, PR 00931-3334, USA
[6]Department of Chemistry and Department of Physics, University of St. Andrews, St. Andrews KY16 ST, United Kingdom

**Abstract**

The out-of-plane and in-plane polarization of $(Pb_{0.6}Li_{0.2}Bi_{0.2})(Zr_{0.2}Ti_{0.8})O_3$(PLBZT) thin film has studied in the dark and under illumination of a weak light source of a comparable bandgap. A highly oriented PLBZT thin film was grown on $LaNiO_3$ (LNO)/$LaAlO_3$(LAO) substrate by pulsed laser deposition system which illustrates well-saturated polarization and its significant enhancement under illumination of light. We have employed two configurations for polarization characterization; first deals with out of plane polarization with single capacitor under investigation, whereas second demonstrates the two capacitors connected in series via the bottom electrode. Two different configurations were illuminated using different energy sources and their effects were studied. The latter configuration shows a significant change in polarization under illumination of light that may provide an extra degree of freedom for device miniaturization. The polarization was also tested using positive-up & negative-down (PUND) measurements which confirm robust polarization and their switching under illumination.

*Corresponding Author: Dr. Ashok Kumar (Email: ashok553@nplindia.org)*

1. **Introduction**

The robust and fatigue free ferroelectric polarization is need of future nonvolatile ferroelectric random access memory (NV-FeRAM) elements due to their fast WRITE ability and long durability. In general metal-ferroelectric-metal (MFM) capacitor with out-of-plane polarization configuration has been used for device fabrication and characterization. However, metal electrode diffusion into the ferroelectric layer, development of dead layers at the interface, depolarization field, and broader time relaxation of polarization are major issues for miniaturization of out-of-plane nano-scale devices [1]. Two capacitors connected in series via conducting bottom electrode and their suitability for device applications for various functional properties has not been widely investigated. In this configuration, illumination of thin film heterostructure by a low energy light source not only generates excess optically active charge carriers across the electrode-electrolyte interface but also significantly affects the polarization over the large open area between two capacitors connected in series. In both conditions, polarization, domains, and domain walls could be easily manipulated optically for destructive READ-out techniques [2,3,4]. In recent years numerous works have been carried out for domains dynamics, domains and domain walls active role in above bandgap photovoltaic effects, conducting domain walls, and complex domain patterns[5,6,7,8]. Still, the microscopic understanding of basic physics and origin of complex domains structure and their behavior under illumination of an energetic light source is limited. The devices based on out-of-plane polarization for NV-FeRAM would be interesting and may provide high-density logic bits. The basic and applied research for polarization (in both configurations) will open a new avenue for optical-electronic sensors, and non-volatile memory elements [9,10,11].

The manipulation of functional properties such as polarization and conductivity under illumination of a light source have been studied which may provide an extra degree of freedom during device miniaturization [12,13,14,15]. The illumination of thin film heterostructure by a weak light source may be utilized for the making various polarization states if polarization is coupled and controlled by external light source. These features may be utilized in the fabrication of next generation opto-memory elements, bulk photovoltaics, optical sensors and optoelectronics materials and devices [16,17,18]. We have also discovered giant enhancement in polarization and reproducible photovoltaic effects in bulk PLBZT ceramics under illumination of weak light [19]. Hu *et al.* have shown that one can manipulate the electro-resistance, polarization, photovoltage, and ferroelectric tunnel junction based memory states under light illumination [20].

The ionic polarization and presence of lone pairs (e.g. $Bi^{3+}$ and $Pb^{2+}$) have been widely accepted as the main cause for the development of polarization in the ferroelectric matrix, however recently it has been observed that electronic polarization may significantly alter the ionic polarization due to the displacement of central cations and 2p electron orbitals for oxygen ions [21]. The illumination of light populates the optically active charge carriers near the conduction band that re-localized the spatial distribution of electrons in the matrix which in turn modulates the polarization vector. Many optically active polar compounds have been studied which are capable of storing optical information [22,23]. The illumination of light on thin film heterostructure can easily modify the trap levels, fill the voids, nullify the free charge carriers near the grains and grain boundaries [24]. Bismuth oxide is well studied ferroelectric which traps light due to low bandgap and optically active charge centers, and it can also change their shape and size under illumination [25] The origin of electronic polarization and their contribution to ionic polarization and total polarization is not well understood. The effects of The illumination of light on ferroelectric polarization, fatigue, retention, and conductivity have been widely studied and observed in several ferroelectric materials [26,27].The control of dielectric and polarization properties by external light source might be used for high-density memory components, ferro-photovoltaics, optoelectronics gadgets, *etc* [10,16].

In this report polarization and electrical properties have been investigated in two different out-of-plane configurations. We have carried out polarization study on ferroelectric thin films with special focus on two out-of-plane capacitors connected in series via bottom electrode since this particular configuration used an enhanced optically active area between two in-plane semi-transparent Au-electrode. In another configuration a standard capacitor with top and bottom electrode was studied. In these conditions, two PLBZT capacitors are connected in series by bottom LNO electrode.

## 2. Experimental Methods

The ceramic target of $(Pb_{0.6}Li_{0.2}Bi_{0.2})(Zr_{0.2}Ti_{0.8})O_3$ (PLBZT) was prepared by a conventional solid-state reaction route and used for fabrication of thin films [19]. Pulsed laser deposition (PLD) technique was employed to grow highly oriented PLBZT films on $LaNiO_3$ (LNO)/$LaAlO_3$(LAO) substrate. Initially, LAO substrate was ultrasonicated in acetone for two minutes and then in methanol for 30 minutes to remove the native oxides and impurities present on the substrate before laser ablation. The thin conducting LNO layer (50 nm) was first deposited on the LAO (100) single crystal substrate for preferential growth and the bottom electrode. The LNO layer was grown at 700 °C temperature, 200 mTorr oxygen partial pressure, 5 Hz frequency, and 1.5-2 J/cm$^3$ energy density, followed by annealing at the same temperature and oxygen ambient of 300 Torr for 30 minutes with slow cooling (5 °C/min) down to room temperature. Later a shadow mask is utilized to expose the bottom electrode for characterization. After mounting the shadow mask, PLD chamber was further taken down to a base pressure of $10^{-6}$ Torr to grow the thin films on LNO coated substrate. The PLBZT thin films of average thickness 150 nm were deposited on this substrate using a KrF Excimer laser (wavelength-248 nm, frequency-5 Hz) with an energy density nearly 1.5 J/cm$^3$. The deposition conditions such as deposition temperature, oxygen partial pressure, laser energy, and frequency for PLBZT thin films was kept at 700 °C, 80 mTorr, 1.5 J/cm$^3$ and 5 Hz, respectively. As-grown films were annealed in-situ in pure oxygen ambiance at 200 Torr and 700 °C for 30 min with slow cooling (5 °C/min) down to room temperature. The phase purity and orientation of these films were examined by X-ray diffraction (XRD); (Bruker AXSD8 Advance X-ray diffractometer) and their thicknesses were determined using a profilometer. The surface topography of the epitaxial thin film was investigated by atomic force microscopy (AFM) (Model: Nanoscope V, Make: Veeco). To study the ferroelectric and electrical characteristics, semi-transparent gold (Au) electrodes of average thickness 40 nm in a square shape having a length 100 μm X 100 μm with a separation of 1000 μm were deposited by using a shadow mask and performed electrical characterization as shown in the schematic diagram (see Fig.3 (a &b)). Ferroelectric properties were measured at room temperature using a Radiant ferroelectric tester. The current-voltage characteristics were performed under illuminated of laser and white light using Keithley-236 source-meter at ambient conditions.

Three conditions are employed to check ferroelectric polarization; *i.e.* dark (D), under illumination of weak white light (W) with spectral range from 450 nm to 1200 nm and illumination of monochromatic laser light (L) having wavelength 405 nm. The intensity of white light and laser source was nearly 60 mW/cm$^2$ and 30 mW/cm$^2$, respectively which illuminated at the center of the semi-transparent electrode for the single capacitor and the center point between two electrodes for the double capacitor. The nomenclatures 1D, 1W, 1L and 2D, 2W, 2L for dark, illumination of white light, and illumination of laser for SC and DC, respectively.

### 3. Results and Discussion

### 3.1 X-ray diffraction (XRD)

Figure. 1(a) shows the XRD patterns of highly oriented PLBZT thin films with an average thickness about 150 nm grown on LNO (50 nm) coated LAO substrate. All peaks are identified and mentioned in the fig. 1 (a), however (#) represents an alien peak of the substrate. The XRD peak at 29.4° and hump XRD peak near 37° represent domain twining and impurities of LAO substrate (not shown here). Sometime it develops due to imperfect growth, cutting and polishing conditions of single crystal substrate. The LNO peak is very weak and also shadowed by the highly intense (002) peak of LAO (marked in the Fig.1 (a)). The XRD patterns of as received LAO substrate is given separately in supplementary information (Suppl. Fig.1). The profilometer data confirming the film thickness (~150 nm) is also shown in supplementary information (Suppl. Fig.2). The close view of PLBZT/LNO/LAO (002) peak with bulk PLBZT XRD peaks is shown in figure 1 (b). The XRD patterns are shown in figure 1 (a) & (b) confirmed phase purity and preferential orientation along (001) plane. A close view of (200) & (002) planes of bulk PLBZT ceramics indicates that the PLBZT film possesses a large tensile strain along (200) plane (in-plane) and compressive strain along (002) plane (out of plane) compared to the bulk lattice. It occurs due to possible large lattice mismatch of bulk tetragonal PLBZT (lattice parameters a=b=3.929Å and c=4.124Å) [28] compared to the LNO coated rhombohedral LAO substrate (a = 3.79 Å, c = 13.11 Å) [29]. To explore the effect of strain on the ferroelectric and conductivity properties, the strain (ε) in the films due to lattice mismatch (or misfit) between bulk and thin film was calculated using equation; $\varepsilon = \frac{a_{bulk} - a_{film}}{a_{bulk}} \times 100$, where, ε is the lattice strain, $a_{bulk}$ is the lattice constant of the bulk, and $a_{film}$ is the lattice constant of the film. If ε > 0, films will experience in-plane tensile strain related to their bulk counterpart. Nearly 2.5% in-plane tensile strain was observed in the film compared to bulk crystal structure. The magnitude of strain is rather large which favors the condition for orientation of polarization along in-plane which predicts the possibility of large in-plane polarization. Figure 2 (a & b) shows surface morphology of PLBZT thin film over an area of 5x5 μm and 2x2 μm on two different regions that provide an average grain size of 10-25 nm and with average surface roughness of 1-3 nm. A well dense wave nature of granular microstructure was seen in most of the investigated area.

### 3.2 Schematic Illustration of circuit diagram

Figure 3(a) & (b) show a schematic circuit diagram which is used to measure the in-plane and out-of-plane polarization and their enhancement under illumination of a light source. Figure 3(a) and (b) represent the DC and SC configuration with an equivalent electrical circuits Fig. 3(c) and Fig. 3(d), respectively where the top square shape Au electrodes of 100 X 100 μm dimension was fabricated using shadow mask with a separation of 1000 μm. The well saturated and concave nature of polarization loops was observed under dark and illumination of light conditions.

### 3.3. Ferroelectric Properties

The ferroelectric polarization *versus* voltage (P-V) hysteresis loops and displacement currents have been investigated to understand intrinsic polarization behavior of the highly oriented film for SC and DC configurations (Figure. 4(a-d)). The DC device with 100 μm X 1000 μm active area between two electrodes (connected through bottom LNO electrode) was fabricated to probe the polarization. P-V loops were symmetrical and saturated, and almost frequency independent below 10 Hz probe frequency as can be seen from Fig. 4(a). A significant enhancement (nearly 100%) in ferroelectric polarization and displacement current were observed under illumination of monochromatic light for DC configuration, whereas a small change (nearly 15 %) under illumination in ferroelectric polarization (at 50 Hz) were observed in the case of SC configuration. It indicates that a large amount of optically active charge carriers contribute significantly for DC connected in series (due to the large active area) rather than SC configuration under illumination. We have done polarization and PUND analysis for various frequencies and pulse widths. The presented data is the best data where the film shows a maximum change in polarization under illumination of light. The enhancement in polarization under illumination of thin film heterostructure by an energy source may be due to the effective contribution of electronic polarization and orientation of ionic polarization in the direction of applied electric field. Matthew Dawber *et. al.* suggested that change in the spatial distribution of *2p* electron orbitals of oxygen ions in perovskite octahedral may lead to the development of electronic polarization [21] Here illumination of light changes the electronic weight of oxygen 2p orbitals which shifts opposite to the direction of central cations ($Ti^{4+}$/$Zr^{4+}$) responsible for ionic polarization ($P_{ion}$), resulting an additional electronic polarization ($P_{ele}$) in the matrix in the direction of ionic polarization [21]. It is also possible that under illumination of light, a large number of electronic charge carriers jump from valance band to conduction band and developed a shift current due to the asymmetry of the electronic band which may significantly affect the displacement current and overall polarization of the matrix [30].

**3.4 PUND analysis**

A PUND (positive-up & negative-down) analysis has been carried out to check the intrinsic polarization in both the configurations. We have used a 1ms (width) pulse signal with 15 V pulse amplitude to measure the polarization of DC connected through bottom electrode and for a SC with out-of-plane configuration 10 ms (width) pulse signal with 10 V pulse amplitude. These pulse widths and pulse amplitudes are optimized for both configurations to achieve maximum control of polarization under illumination. Figure 5 (a,c) shows PUND plots for various switchable and non-switchable polarization (remanent and saturation) of PLBZT film in SC and DC modes of configuration. The net switchable polarization (dP) is as follows: $dP = P^* - P^\wedge$ = where $P^* $ = (switchable polarization + non-switchable polarization) and $P^\wedge$ = (non-switchable polarization); $dP_r = P^*_r - P^\wedge_r$ where $P^*_r$ = (switchable remanent polarization + non-switchable remanent polarization) and $P^\wedge_r$ = (non-switchable remanent polarization). The PUND analysis indicates nearly 10% enhancement in switchable polarization compared to the 100% enhancement in total polarization from the out-of-plane (DC) P-V hysteresis loops under illumination of light. These results suggest that nearly 10% enhancement in intrinsic remanent polarization under illumination, a similar magnitude of enhancement can be seen for the SC configuration. The large difference between the switching and non-switching responses and significant enhancement under illumination suggest a possible intrinsic polarization with and without light and may be useful for developments of various logic states.

To measure the lifetime of the ferroelectric device under application of the external electric field, a PUND based fatigue characteristics has been carried out on both configurations. We have applied a train of a pulse having a pulse width 1 ms and amplitude 10 V across the capacitor to check the fatigue characteristic for SC and DC configuration. The switchable remanent polarization *versus* cumulative time under dark and illumination of white light is displayed in Figure. 5(b,d). The fatigue results suggest nearly 30% degradation of remanent polarization (2dP) after $10^5$ cycles under white light illumination conditions for both the configurations, however, nearly unchanged polarization under dark conditions. These results suggest that the photo-generated charge carriers which relocate the electronic weight in the lattice and saturate after a long period, however, the polarization under illumination are still higher than the dark condition.

### 3.5 Band-gap calculation

Figure 6 shows the Tau plot of PLBZT thin films which indicate direct bandgap (~ 3.1 eV) where almost 75% transmission impede however another slope can be seen where almost 100% transmission stop. The last slope gives a direct band nearly 2.5 eV which may be due to the presence of defects in the system. These values are less compared to the basic Ti-rich PZT films (~ 3.5 eV).

### 3.6 Conduction mechanism in ferroelectric thin films

Figure 7 (a,c) shows the leakage current behavior in the dark and under illumination of an external light source for SC and DC configurations, respectively. It is almost symmetrical in nature under both positive and negative bias till ± 30 V and ± 15 V for DC and SC configurations, respectively. The average dark current is about ~$10^{-7}$ A at ±30 V and $10^{-5}$ A at ± 15 V for DC and SC, respectively. An order of higher current magnitude is observed in two capacitors connected in the series configuration under illumination of white light and monochromatic light. This enhancement in leakage current is due to optically active charge carriers which gain energy under light illumination, jump from valence band to conduction band, modify the Fermi level and cross the Schottky electrode-ferroelectric barrier height. The optically active charge carriers and cations trap the light and modify the surface charge layer, redistribute the charge carriers across grain boundaries, and neutralization of depolarization field at electrode-film interface. In DC configuration with large distance (1000 μm) between the electrodes to avoid the contribution of photocurrent in intrinsic polarization, but at the same time illumination of light also helps to develop electronic polarization which improves ionic polarization and orient the in-plane polarization vector along the electric field. In general, the photo-generated charge carriers annihilate within few micrometer distances, but some of them hop to the nearest neighbor site and significantly contribute the electrical conduction.

All possible conduction mechanisms have been employed to understand the conduction mechanism of PLBZT thin films, but none of them meet their theoretical criteria except trap assisted space charge limited conduction (SCLC) mechanisms [31]. Figure 7 (b,d) shows the linear fitting of the current-voltage data in three different voltage regions, low voltage region follows Ohmic behavior, intermediate voltage meets the basic criteria of trap-assisted current conduction process, and high voltage regions deep level trap assisted current conduction with an exponent higher than 3. In the intermediate and high voltage regions, the injected charge carrier increases very rapidly and quickly filled the trap centers and developed a condition of trap-free conduction with a very narrow window of applied electric field. At low applied voltage, I-V characteristics followed Ohmic current behavior because the densities of thermally generated free carriers inside the films were larger than the injected charge carriers. When the internal field is dominated by space charge (either from free or trapped carriers), the current (I) is expected to follow a power-law dependence on voltage (I α $V^n$) [33]. I-V characteristics show that at a higher voltage the slope is *n>3* it may be due to more generation of space charges due to the bottom conductive electrode. Room temperature slope exponents ($n$) were found 1, 2 and 3 which represent initial Ohmic conduction, intermediate trap-assisted conduction and high voltage deep level trap-free conduction, respectively which suggest that SCLC conduction is likely to be the dominant leakage mechanism.

The present investigation was intended to check the control of polarization by various weak light sources to develop multistate NV-FeRAM logic elements. In this regards, we have characterized two basic configurations of out of plane capacitor-type device. First, we probed a conventional single capacitor (SC) with semi-transparent top Au electrode and conducting LNO bottom electrode. In this particular configuration, the device provides nearly 12% enhancements in polarization under illumination of a weak light source. To improve the effect of light on polarization, we further studied an alternative device structure where two single capacitors situated nearby were connected through bottom electrode in series. It was expected that the second configuration should provide significant enhancement in polarization under illumination of a weak light source since this particular configuration take advantage of the surface polarization as well as out of plane polarization and its maximum interaction with light (surface area). As expected nearly 100% enhancement in polarization was observed for DC configuration for a weak light source. The DC configuration will also provide an extra degree of freedom for miniaturization of an optically active device for future microelectronics.

## 4. Conclusions:

We have investigated two out-of-plane configurations of PLBZT device; (i) the DC connected through conducting bottom LNO electrode and (ii) a SC with top Au electrode and bottom LNO, respectively. Both out-of-plane configurations of the device were tested for possible ferroelectric polarization-based logic elements. A detailed investigation of well-saturated hysteresis loops, enhancement of polarization under illumination of a light source, PUND based switchable polarization, and displacement current suggests the intrinsic nature of polarization and displacement current and its enhancement under illumination of light. The substitution of optically active cations significantly reduced the direct bandgap (~ 3.1 eV) compared to the basic Ti-rich PZT films. Almost thirty percent polarization fatigue was observed for a large number of test cycles under illumination of light. However, negligible reduction in polarization was seen in dark conditions. The current conduction process meets the criteria of space charge limited mechanism which becomes trap-free at high applied voltage. These results may help the engineers to design out-of-plane ferroelectric memory cell with electrically write and optically read process.

**Acknowledgment**

Ashok Kumar acknowledges the CSIR-NPL, India project PSC-0111. Hitesh Borkar would like to acknowledge the UGC-SRF to provide fellowship to carry out the Ph.D program. Authors sincerely thank the Director, NPL, Dr. Ranjana Mehrotra and Dr. Sanjay Yadav for their constant encouragement.

**Figure captions**

**Fig. 1 (a)** X-ray diffraction patterns of PLBZT/LNO/LAO epitaxial thin film over a wide range of Bragg's angle. **Fig. 1(b)** shows the close view of (002) PLBZT film plane and compared with lattice plane of bulk PLBZT ceramics in the same Braggs angle range.

**Fig. 2** Surface topography image of PLBZT thin film over an area of (a) 5μm x 5μm and (b) 2μm x 2μm at two different places on the surface.

**Fig.3** (a) shows a schematic illustration of the testbed device structure for two out-of-plane capacitors *i.e.* double capacitor (DC) connected through bottom electrode, (b) single capacitor (SC) device structure, (c) circuit diagram for DC in series connection and (d) circuit diagram of SC for out-of-plane configuration.

**Fig.4** (a) illustrates ferroelectric polarization and (b) switching of displacement under three different conditions i.e. 2D, 2W, & 2L for DC at 10 Hz, respectively. Fig. 3 (c) and (d) illustrate ferroelectric polarization and switching of displacement current under three different conditions i.e. 1D, 1W, & 1L for SC at 50 Hz, respectively.

**Fig.5** (a) & (b) show the PUND data and fatigue characteristics of DC that utilized a pulse signal of width 1 ms and amplitude 15 V for PUND and 10 V for fatigue, respectively. Fig 5 (c) & (d) show the PUND data and fatigue characteristics of SC that utilized a pulse signal of width 10 ms and amplitude 10 V, respectively.

**Fig.6** Illustrates the Tau plot of PLBZT thin film without bottom electrode to calculate the bandgap of the system.

**Fig.7** (a) & (b) show the leakage current (I) as a function of applied voltage (V) and linear fitting of I-V characteristics in log-log scale to check the suitability of SCLC mechanism for DC mode under 2D, 2W, and 2L, respectively. Fig. 6 (c) and (d) show the leakage current (I) as a function of applied voltage (V) and its linear fitting for SC mode under 1D and 1W condition, respectively.

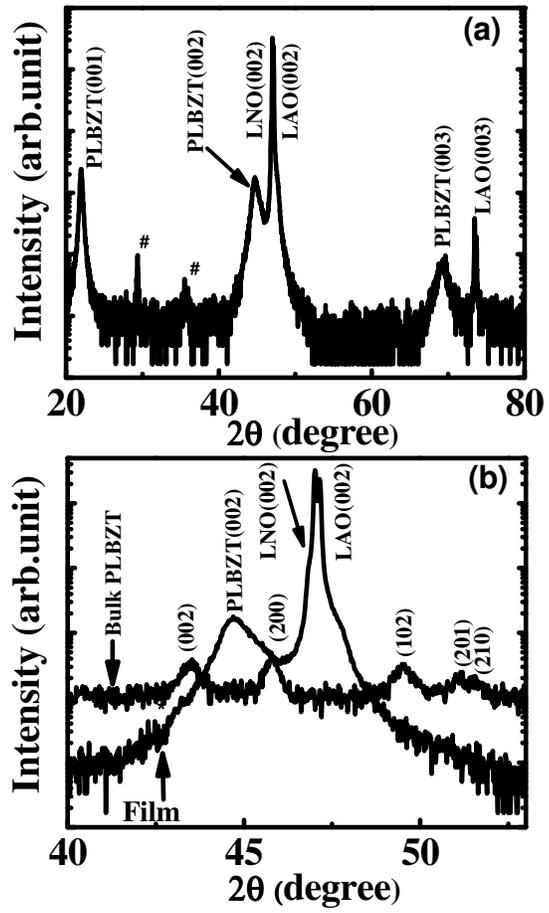

**Figure 1**

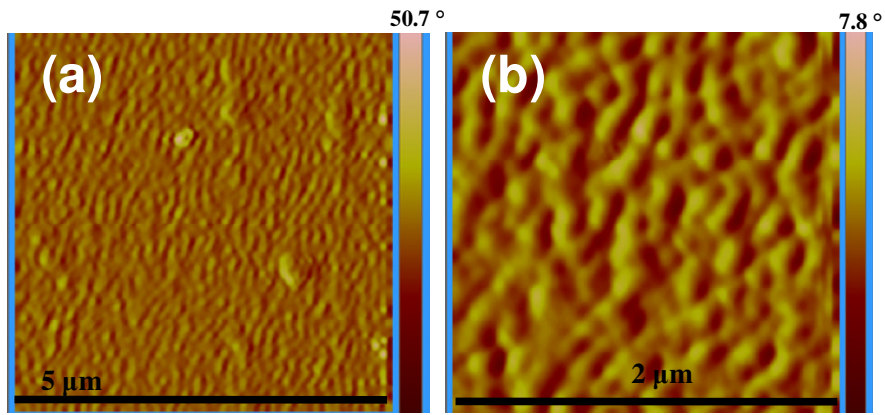

**Figure 2**

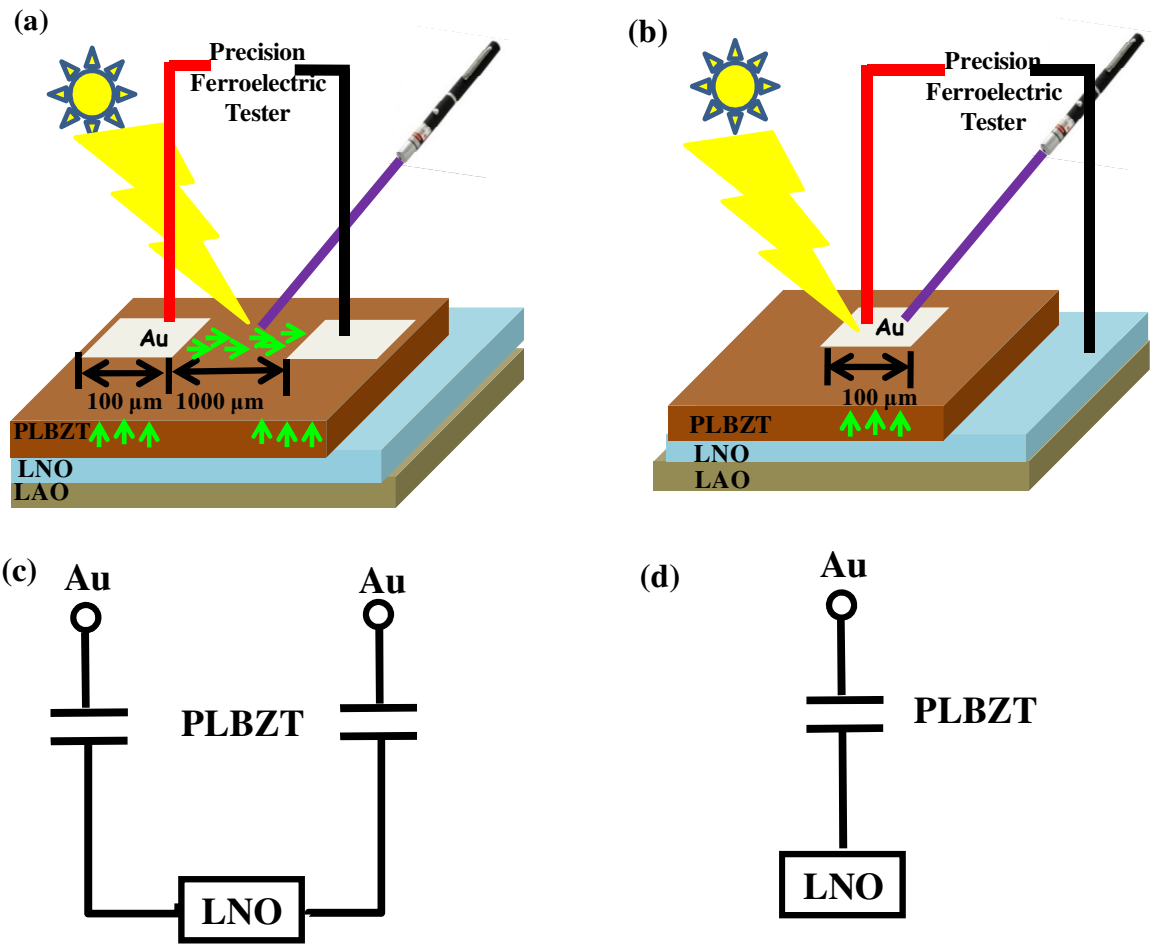

**Figure 3**

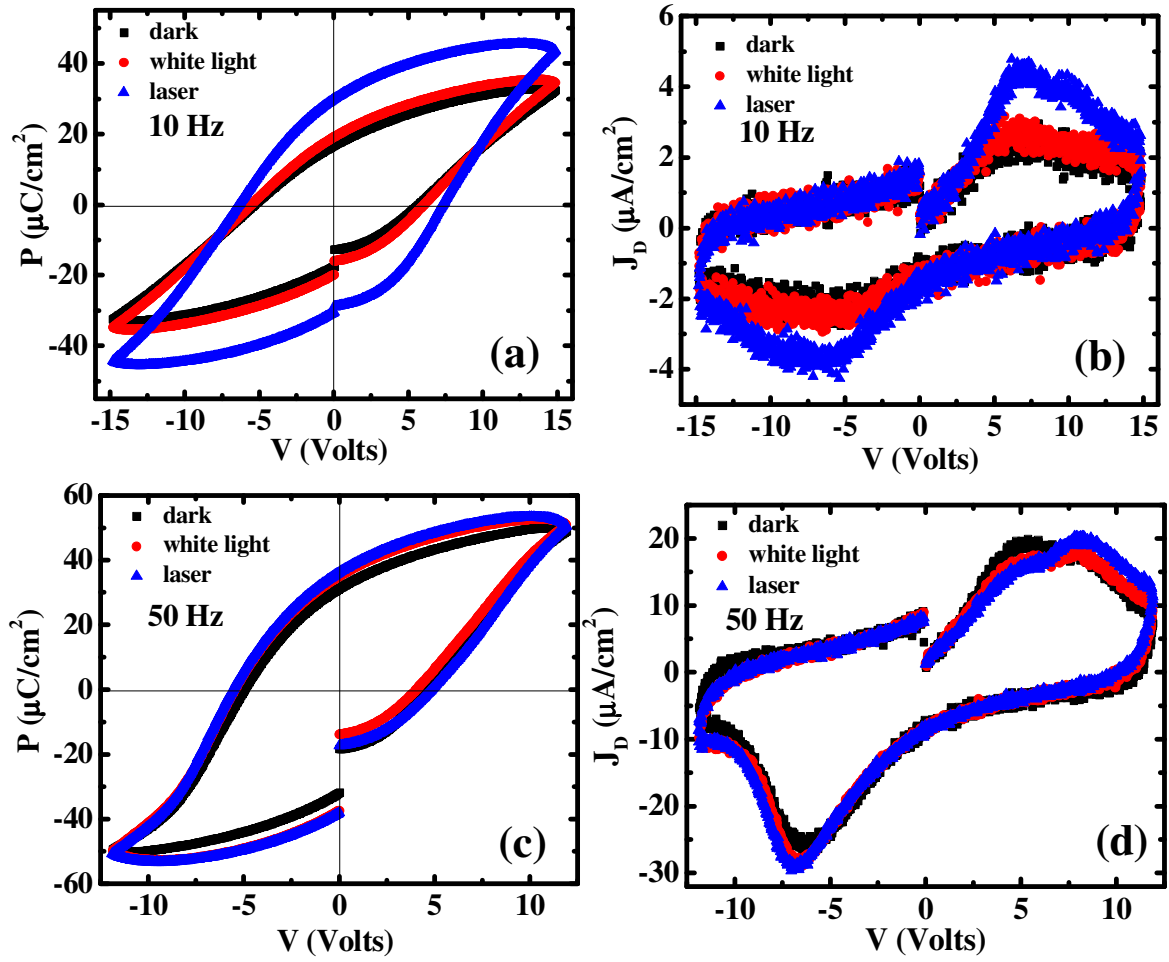

**Figure 4**

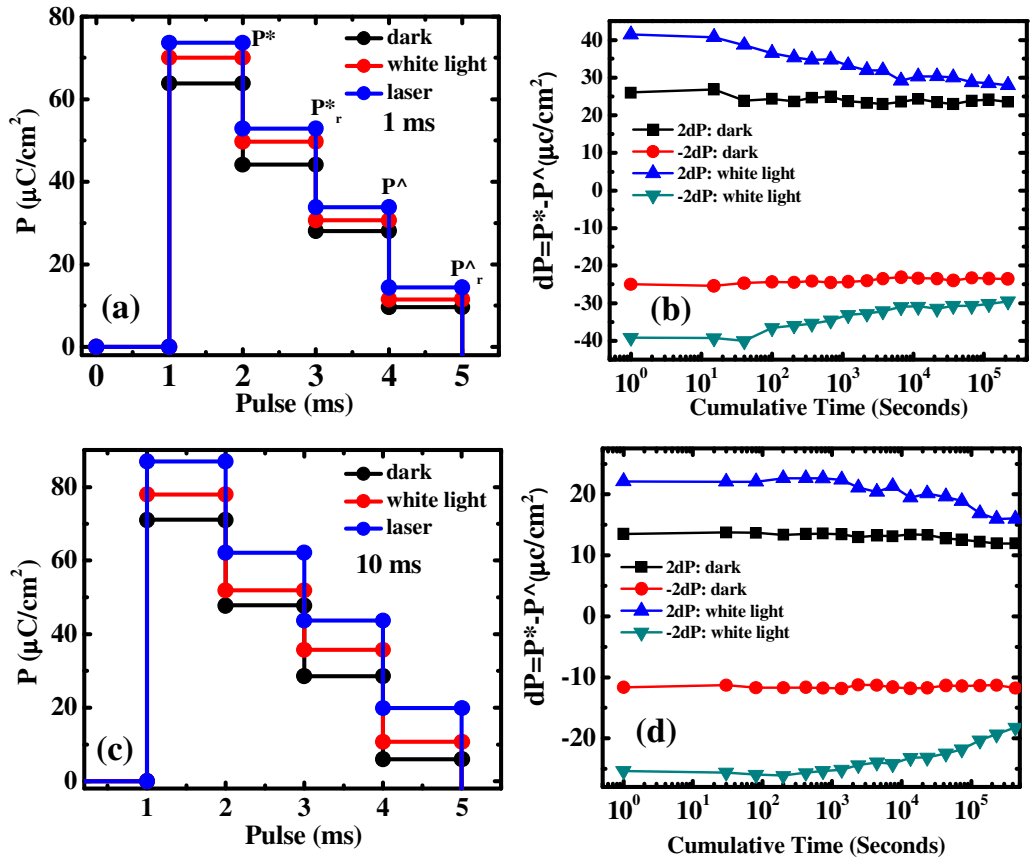

**Figure 5**

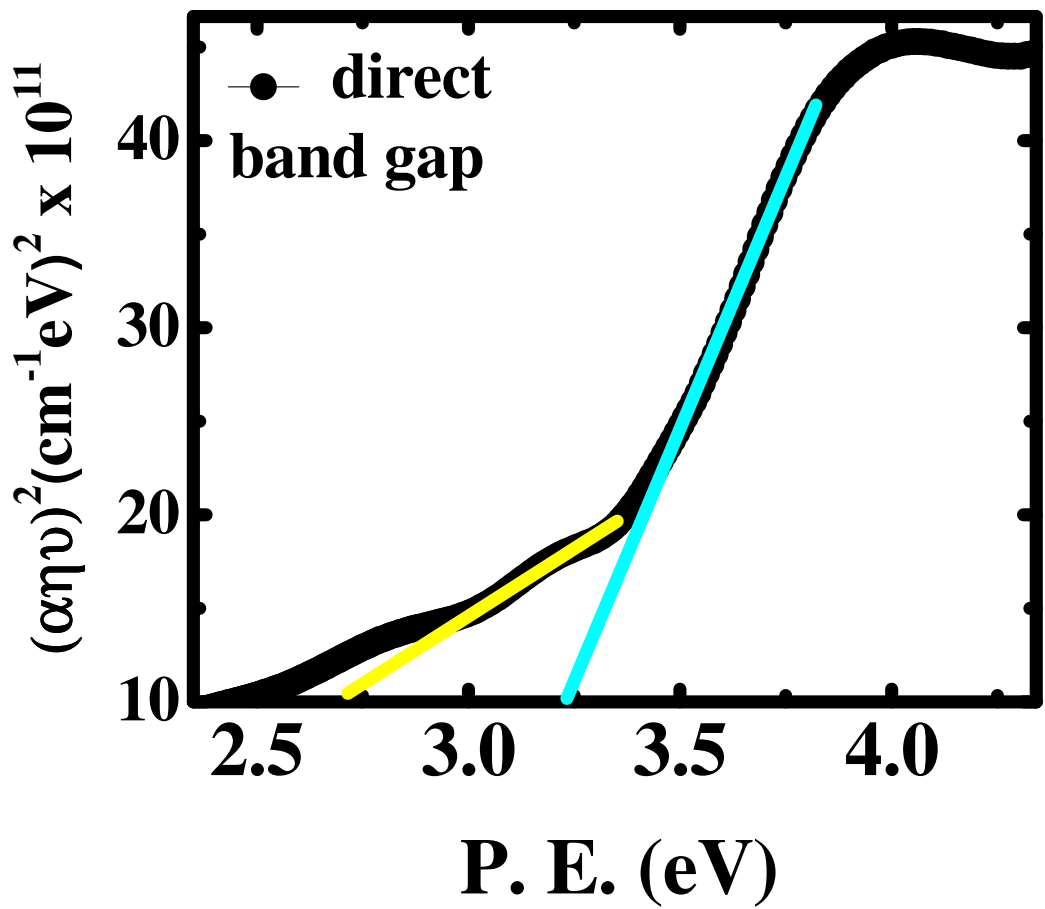

Figure 6
21

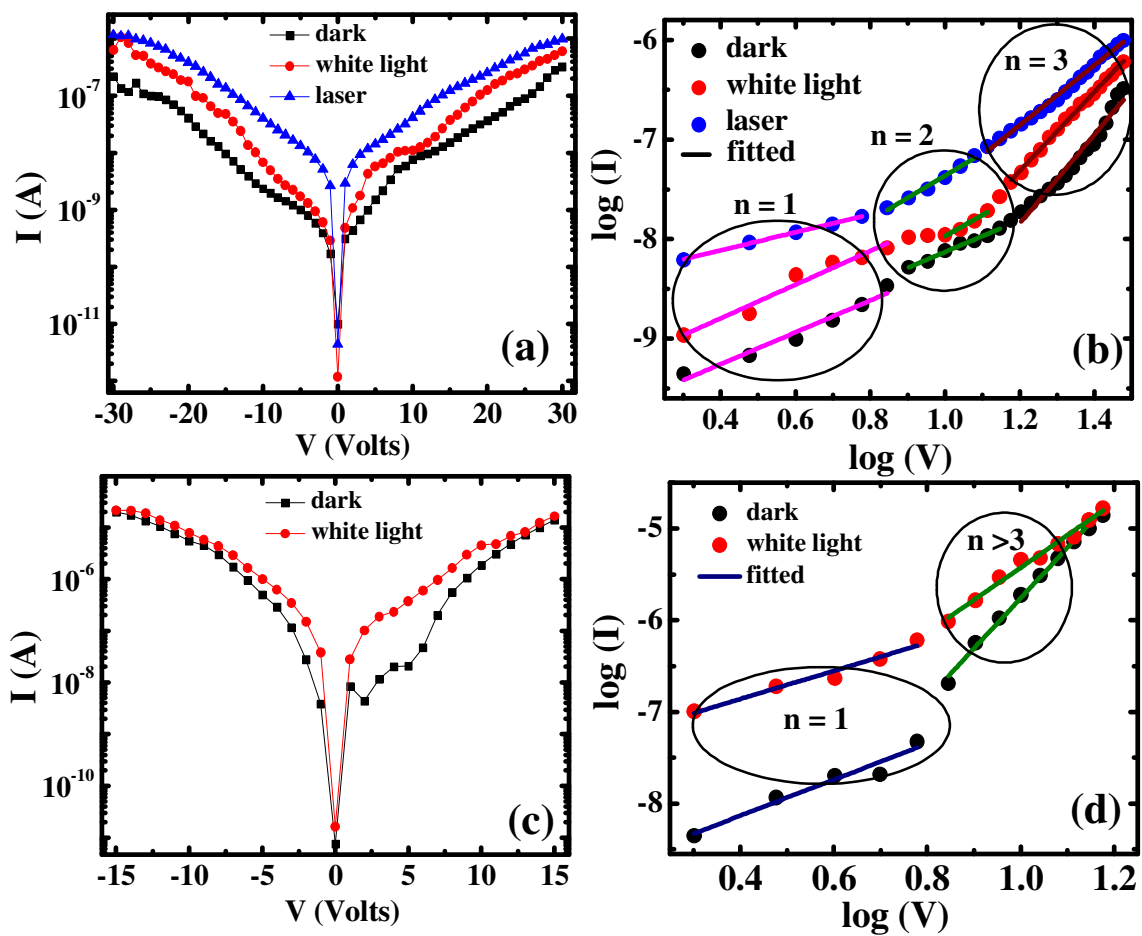

**Figure 7**

# References:

[1] Dawber M., Rabe K. M., Scott J. F., 2005 *Rev. Mod. Phys.* **77**, 1083.
[2] Catalan G., Seidel J., Ramesh R., Scott J. F., 2012, *Rev. Mod. Phys.* **84**, 119.
[3] Matzen S., Nesterov O., Rispens G., Heuver J. A., Biegalski M., Christen H. M., Noheda B., 2014, *Nature Communications* **5**, 4415.
[4] Tsymbal E. Y., Gruverman A., Garcia V., Bibes M., Barthelemy A., 2012 *MRS Bull.* **37**, 138.
[5] Fu H., Bellaiche L., 2003 *Phys. Rev. Lett.* **91**, 257601.
[6] Schilling A., Byrne D., Catalan G., Webber K. G., Genenko Y. A., Wu G. S., Scott J. F. and Gregg J. M., 2009, *Nano Lett.* **9**, 3359.
[7] Balke N., Winchester B., Ren W., Chu Y. H., Morozovska A. N., Eliseev E. A., Huijben M., Vasudevan R. K., Maksymovych P., Britson J., *et. al*, 2012 *Nat. Phys.* **8**, 81.
[8] Yang S. Y., Seidel J., Byrnes S. J., Shafer P., Yang C. H., Rossell M. D., Yu P., Chu Y. H., Scott J. F., Ager J. W., III, Martin L. W., Ramesh R., 2010 *Nature Nanotechnology* **5**, 143.
[9] Uchino K., 2000 *Ferroelectric Devices*, Marcel Dekker, New York.
[10] Lines M. E. and Glass A. M., 1977 *Principles and Applications of Ferroelectrics and Related Materials*, Oxford University, New York.
[11] Scott J. F., 2007, *Science* **315**, 954.
[12] Kholkin A. L. and Setter N., 1997 *Appl. Phys. Lett.*, **71**, 2854.
[13] Polkawich R. L. and Trolier-McKinstry S., 2000 *J. Mater. Res.*, **15**, 2505.
[14] Kholkin A., Boiarkine O., and Setter N., 1998 *Appl. Phys. Lett.* **72**, 130.
[15] Zubko P. and Triscone J. M., 2009 Nature, **460**, 45.
[16] Wadhawan V. K., Introduction to Ferroic Materials, Gordon and Breach Science Publishers, Amsterdam (2000).
[17] Alexe M. and Hesse D., 2011 *Nat. Commun.* **2**, 256.
[18] Choi T., Lee S., Choi Y. J., Kiryukhin V. and Cheong S. W., 2009, *Science*, **324,** 63.
[19] Kumar A., Borkar H., Rao V., Tomar M., Gupta V., Multi-States Nonvolatile Opto-Ferroelectric Memory 2016, *Indian patent No* #: 201611001338.
[20] Hu W. J., Wang Z., Yu W., Wu T., *Nature Comm.* **7**, 10808 (2016).
[21] Dawber M. 2012 *Physics* **5**, 63.
[22] Land C. E. and Smith W. D., 1973 *Appl. Phys. Lett.*, **23**, 57.
[23] Maldonado R., Fraser D. B., and Meitzler A. H., 1975 *In Advance in image Pickup and Display, edited by B. Kazan Academic*, New York, **65**, 168.
[24] Borkar H., Tomar M., Gupta V., Scott J. F., Kumar A., 2015 *Appl. Phys. Lett.*, **107(12)**, 122904.
[25] Kundys B., Viret M., Colson D., and Kundys D. O., 2010 *Nat. Mater.* **9**, 803.
[26] Polkawich R. L. and Trolier-McKinstry S., 2000 *J. Mater. Res.* **15**, 2505.
[27] Kholkin A., Boiarkine O., and Setter N., 1998 *Appl. Phys. Lett.* **72**, 130.
[28] Borkar H., Rao V., Tomar M., Gupta V., Scott J. F., Kumar A., 2017 *RSC Advances* **7**,12842 .
[29] Luo X. and Wang B., 2008 *J. Appl. Phys.* **104**, 073518.
[30] Young S. M., Rappe A. M., 2012, *Phys. Rev. Lett.* **109**, 116601.
[31] Tredgold R. H., 1965 *Space Charge Conduction in Solids*, Elsevier Publishing Company, Amsterdam.